# A Modular, Low-Cost IoT System for Environmental and Behavioural Monitoring in Cultural Heritage Sites

## Juan Palomeque-Gonzalez

**Institute of Evolution in Africa (IDEA), Alcalá University, Covarrubias 36, 28010, Madrid, Spain**

jfpalomeque.gonzalez@gmail.com

https://orcid.org/0000-0002-1895-895X

# Introduction

Cultural heritage, encompassing historical structures, artefacts, and traditions, forms an irreplaceable pillar of human identity and socio-economic development (Zombory 2022; Donders 2020). UNESCO recognises its preservation as integral to human rights, particularly as heritage sites face escalating threats from climate change, mass tourism, and resource limitations (Donders 2020; Haugen et al. 2018). For instance, rising global temperatures and humidity fluctuations directly endanger fragile artefacts and masonry, while unchecked visitor behaviour accelerates structural degradation (Camuffo, Della Valle, and Becherini 2022; Lee and Lee 2019). Concurrently, heritage tourism drives economic growth, contributing significantly to local communities in regions such as Sarajevo and Lisbon (Meha, Tahiri, and Zhubi 2020; Silva, Coelho, and Henriques 2020). However, this paradox—balancing accessibility with preservation—demands innovative solutions to monitor environmental variables and human activity proactively.

Existing environmental monitoring systems for heritage conservation often rely on costly, proprietary hardware, limiting their scalability and adaptability (Manfriani et al. 2021; Bacco et al. 2020). Commercial solutions, while effective, suffer from inflexibility, vendor lock-in risks, and prohibitive costs for resource-constrained institutions (Leccese et al. 2014; Mitro, Krommyda, and Amditis 2022). For example, precision sensors deployed in Genoese museums lack interoperability, hindering system upgrades (Manfriani et al. 2021). Conversely, open-source initiatives, such as Cordoba's low-cost temperature monitoring network, demonstrate potential but omit critical IoT features like real-time cloud integration (Mesas-Carrascosa et al. 2016). This gap between affordability and functionality leaves many heritage sites, particularly in developing regions, without viable tools for preventive conservation (Sudantha et al. 2018).

This study addresses these challenges by designing and evaluating a modular, low-cost IoT system tailored for heritage conservation. The proposed system integrates a Wireless Sensor Network (WSN), edge computing, and cloud analytics to monitor environmental variables—such as temperature, humidity, and visitor proximity—while employing computer vision to detect risky behaviours. Built on off-the-shelf components like ESP32 microcontrollers and Raspberry Pi, the framework prioritises cost-efficiency (<£200 per node) and adaptability (Silva, Coelho, and Henriques 2020; Krommyda, Mitro, and Amditis 2022). For instance, Bluetooth Low Energy (BLE) enables plug-and-play sensor integration, while JSON configuration files allow customisation without firmware changes (Dian, Yousefi, and Lim 2018; Vidovszky and Pintér 2022).

The system's contributions are threefold. First, it advances modularity through open-source hardware-software co-design, permitting seamless upgrades—such as swapping DHT22 sensors for air quality monitors—to meet evolving conservation needs (Agbota et al. 2014). Second, it combines edge-cloud capabilities: a Raspberry Pi hub locally analyses data (e.g., classifying visitor behaviour via fine-tuned Vision Transformers) while Azure Cloud ensures secure, scalable storage

(Shi et al. 2016; Zhang et al. 2021). Third, the proof-of-concept validation in a simulated museum environment demonstrates practical feasibility, achieving 95% accuracy in image-based behaviour recognition and reliable real-time reporting (Casillo et al. 2022; Piccialli et al. 2024).

By bridging the gap between low-cost DIY systems and industrial IoT, this work empowers heritage custodians—from local museums to UNESCO World Heritage Sites—to adopt proactive preservation strategies without budgetary overreach. The following sections detail the system's architecture, experimental validation, and implications for sustainable heritage management.

## Research Aim

This study aims to design, implement, and evaluate a low-cost, modular Internet of Things (IoT) system for monitoring environmental conditions and visitor behaviour in cultural heritage settings. Existing solutions often rely on proprietary, high-cost hardware, limiting accessibility for resource-constrained institutions and hindering adaptability to diverse conservation needs (Manfriani et al. 2021; Silva, Coelho, and Henriques 2020). To address these gaps, the proposed system integrates off-the-shelf components—including ESP32 microcontrollers, Raspberry Pi edge computing, and Azure cloud services—to create an open-source framework prioritising flexibility, scalability, and cost-efficiency. Key objectives include: (1) developing a wireless sensor network (WSN) capable of measuring temperature, humidity, sound, and proximity using Bluetooth Low Energy (BLE) and Wi-Fi communication; (2) implementing edge-computing strategies for real-time data aggregation and deep learning-based image analysis; and (3) validating the system's performance through deployment in a simulated heritage environment. By combining modular hardware design, edge-cloud integration, and proof-of-concept testing, this research contributes a reproducible solution that empowers heritage custodians to adopt proactive conservation strategies. The system's open-source architecture and low-cost components (<£200 per node) democratise access to advanced monitoring tools, particularly for developing regions, while its modularity ensures adaptability to evolving conservation challenges, from climate resilience to visitor management (Sudantha et al. 2018; Vidovszky and Pintér 2022).

# Materials and Methods

## Related Work

The application of IoT technologies in cultural heritage conservation has gained traction over the past decade, with numerous studies exploring environmental monitoring systems for historical buildings and museum collections (Table 1). Early initiatives, such as Agbota et al. (2014), demonstrated the feasibility of wireless sensor networks (WSNs) for remote environmental assessment, employing air quality sensors and cloud-based data storage. Similarly, Grassini and her team (Grassini et al. 2017) developed a cloud-enabled system for museum microclimate monitoring, integrating temperature and humidity sensors with real-time visualisation tools. These projects laid the groundwork for IoT applications in heritage conservation but often relied on proprietary hardware, limiting scalability and adaptability (Manfriani et al. 2021).

Recent advancements have seen a shift towards low-cost, open-source solutions. For instance, Mesas-Carrascosa's team (Mesas-Carrascosa et al. 2016) deployed Arduino-based sensors to monitor temperature and humidity in Cordoba's Mosque-Cathedral, achieving significant cost savings compared to commercial systems. Similarly, Silva's group (Silva, Coelho, and Henriques 2020) implemented a low-cost data-logging system in Lisbon's Jerónimos Monastery, demonstrating the potential of DIY approaches for heritage monitoring. However, these systems often lack advanced IoT features, such as real-time cloud integration or edge computing capabilities, limiting their utility for proactive conservation (Vidovszky and Pintér 2022).

Commercial systems, while robust, present several limitations. High-cost precision sensors, such as those used in Genoa's museum collections, offer unparalleled accuracy but are prohibitively expensive for widespread adoption (Manfriani et al. 2021). Additionally, proprietary hardware often lacks interoperability, making system upgrades or expansions challenging (Leccese et al. 2014). For example, the Colosseum's permanent wireless monitoring system, while effective, relies on specialised vibration sensors and lacks modularity (Monti et al. 2018). These constraints highlight the need for flexible, cost-effective solutions that balance performance with accessibility.

Emerging trends in IoT for heritage conservation include the integration of deep learning and computer vision. Casillo and his team (Casillo et al. 2022) employed convolutional neural networks (CNNs) to detect structural damage in Pompeii, while Piccialli's team (Piccialli et al. 2024) used machine learning to analyse visitor behaviour in museums. These approaches demonstrate the potential of AI-enhanced monitoring but often require significant computational resources, limiting their applicability in resource-constrained settings (Zhang et al. 2021).

In summary, while existing systems have made significant strides in heritage monitoring, gaps remain in terms of cost, modularity, and advanced functionality. This study addresses these limitations by proposing a low-cost, modular IoT system that integrates edge computing, cloud analytics, and open-source hardware to deliver a scalable solution for heritage conservation.

## System Design and Implementation

### System Architecture

The proposed IoT system comprises three main components: a Wireless Sensor Network (WSN), a computing hub, and a cloud backend. The WSN, built around ESP32 microcontrollers, collects environmental data using a variety of low-cost sensors, including DHT22 (temperature and humidity), HW-416-B (PIR motion detection), E3F-R2N1 (photoelectric switch), and HC-SR04 (ultrasonic distance sensor). These nodes communicate via Bluetooth Low Energy (BLE) or Wi-Fi, depending on data volume and latency requirements (Dian, Yousefi, and Lim 2018). For instance, temperature and humidity readings are transmitted via BLE, while camera nodes use Wi-Fi for image transfer.

The computing hub, a Raspberry Pi 4 Model B, serves as the system's central processing unit. It aggregates data from the WSN, performs edge computing tasks—such as image analysis using fine-tuned Vision Transformers (ViT)—and uploads processed data to the cloud backend. The hub also hosts a local Flask-based dashboard for real-time monitoring, displaying environmental variables and visitor behaviour classifications (Chen and Ran 2019).

The cloud backend, built on Microsoft Azure, provides scalable storage and advanced analytics. Data from the WSN is stored in Azure Blob Storage, parsed using Azure Data Factory, and uploaded to an SQL database for querying and visualisation. A Streamlit-based web application renders interactive dashboards, enabling remote monitoring and analysis (Mansouri and Buyya 2020).

### Modularity

A key design principle of the system is modularity, achieved through hardware-software co-design. The WSN nodes are configurable via JSON files, allowing users to add or remove sensors without modifying firmware. For example, the wsn_config.json file defines sensor types, communication protocols, and data collection schedules, enabling rapid adaptation to different monitoring scenarios (Vidovszky and Pintér 2022). Similarly, the computing hub's Python-based software architecture supports dynamic task allocation, with independent routines for data compilation, image analysis, and cloud uploading.

Low-Cost Design

The system prioritises affordability, with each WSN node costing approximately £20–£30, compared to £100+ for commercial alternatives. For instance, the ESP32 microcontroller (£10) and DHT22 sensor (£5) offer comparable performance to high-cost industrial sensors, such as those used in Genoa's museum collections (Manfriani et al. 2021). The Raspberry Pi hub (£50) and Azure's free-tier services further reduce operational costs, making the system accessible to resource-constrained institutions (Sudantha et al. 2018). A diagram of the WSN and Computer Hub can be find in figure 1

**Ethical and Security Considerations**

Privacy and security are critical considerations, particularly for camera nodes. Images captured by ESP32-CAM modules are processed locally to classify visitor behaviour, with raw data deleted immediately after analysis. Additionally, the system employs HTTPS for secure data transmission and Azure's built-in encryption for cloud storage, ensuring compliance with data protection regulations (Kumar et al. 2020).

# Results

## Deployment and Data Collection

The IoT system was deployed in a simulated museum environment, replicating a heritage site with an exhibition cabinet and visitor interaction zones (figures 2, 3, 4, 5, 6). The setup included three ESP32-CAM camera nodes, a temperature and humidity sensor (DHT22), a PIR motion sensor (HW-416-B), a photoelectric switch (E3F-R2N1), an ultrasonic distance sensor (HC-SR04), and a sound sensor (KY-038). The Raspberry Pi 4 computing hub coordinated data collection and analysis, while Azure Cloud services handled data storage and visualisation.

Over a 24-hour testing period, the system recorded 14,149 data points, comprising sensor readings, trigger events, and image captures. The majority of events (10,636) were routine sensor checks, while 199 triggers were logged for motion, sound, and photoelectric beam interruptions. Additionally, 769 sensor readings and 526 network requests were recorded, alongside 199 images captured by the camera nodes.

## Performance Metrics

### Sensor Accuracy and Power Consumption

The WSN nodes demonstrated reliable performance, with minimal latency and low power consumption (table 2). The DHT22 temperature and humidity sensor recorded readings every 5 minutes, consuming an average of 84 mAh per hour. The PIR motion sensor, triggered by human presence, consumed 114 mAh per hour, while the photoelectric switch and sound sensor used 54 mAh and 60 mAh, respectively. The ultrasonic distance sensor, which transmitted data via BLE advertisements, consumed 54 mAh per hour and achieved a communication latency of 6 seconds per reading.

The ESP32-CAM camera nodes, operating over Wi-Fi, consumed between 120 mAh (idle) and 126 mAh (active) per hour. Image capture and transmission took 4 seconds per event, with no data loss observed during testing.

### Deep Learning Model Performance

Three deep learning strategies were evaluated for behaviour classification using a dataset of 750 images: bespoke artificial neural networks (ANNs), fine-tuned Vision Transformers (ViT), and zero-shot classification with CLIP (table 3). The fine-tuned ViT model achieved the highest accuracy (95% for multimodal classification and 100% for binary classification), outperforming bespoke

CNNs (93%) and RNNs (69%). However, the ViT model required the longest training time (14 minutes 47 seconds for multimodal classification).

The zero-shot CLIP model, while requiring no training, achieved lower accuracy (64% for binary classification). Notably, it performed well in identifying negative behaviours (69% accuracy) but struggled with positive behaviours (51% accuracy). This highlights the potential of zero-shot approaches for rapid deployment, albeit with limitations in precision.

## Cloud Infrastructure Reliability

The cloud backend demonstrated robust performance, with a 100% success rate for data uploads and processing. Azure Blob Storage handled unstructured data (e.g., images and CSV files), while Azure Data Factory parsed and stored records in an SQL database. The Streamlit dashboard provided real-time visualisation of environmental variables and event logs, accessible remotely via a web interface.

Cost analysis revealed that the cloud infrastructure could operate at an estimated £180 per month under full deployment conditions (figure 7). However, the use of Azure's free-tier services during testing ensured zero operational costs, making the system financially viable for small-scale implementations.

# Discussion

The results demonstrate the feasibility of the proposed IoT system as a modular, low-cost solution for heritage conservation. The system's ability to reliably monitor environmental variables and visitor behaviour in a simulated museum environment underscores its potential for real-world deployment. The use of off-the-shelf components, such as ESP32 microcontrollers and Raspberry Pi, ensures affordability (<£200 per node) without compromising functionality. For instance, the DHT22 sensor achieved comparable accuracy to high-cost alternatives, while the photoelectric switch provided 100% reliability in detecting unauthorised intrusions. This balance of cost and performance makes the system accessible to resource-constrained institutions, particularly in developing regions where heritage conservation is often underfunded (Sudantha et al. 2018).

A key advantage of the system is its modularity, which allows for easy customisation and scalability. Unlike proprietary solutions, which often require costly upgrades or replacements, the proposed framework supports plug-and-play integration of new sensors and modules. For example, the JSON-based configuration file (wsn_config.json) enables users to add or remove nodes without modifying firmware, reducing maintenance overhead and extending the system's lifespan. This flexibility is particularly valuable for heritage sites with evolving monitoring needs, such as those adapting to climate change or increasing visitor numbers (Haugen et al. 2018).

The system's edge-cloud architecture further enhances its practicality. By performing data aggregation and image analysis locally on the Raspberry Pi hub, the system minimises latency and reduces dependency on cloud services. This is especially beneficial for remote or off-grid locations with limited internet connectivity. At the same time, Azure Cloud integration ensures secure, scalable storage and remote access to monitoring data, enabling heritage custodians to make informed decisions in real time. The Streamlit dashboard, for instance, provides an intuitive interface for visualising environmental trends and identifying potential risks, such as sudden humidity spikes or visitor misconduct.

Despite these strengths, the system has several limitations. First, its reliance on Wi-Fi for camera node communication restricts deployment in areas without reliable internet infrastructure. While BLE offers a low-power alternative for sensor nodes, its limited range (1–10 metres) may pose challenges in large or complex heritage sites. Second, the system's power consumption, though low compared to commercial alternatives, remains a concern for long-term, battery-operated deployments. For example, the ESP32-CAM nodes consume up to 126 mAh per hour, necessitating frequent battery replacements or alternative power sources, such as solar panels. Finally, the small-scale testing environment, while sufficient for proof-of-concept validation, does not fully replicate the challenges of real-world deployment, such as extreme weather conditions or vandalism.

The implications of this work extend beyond technical innovation. By democratising access to advanced monitoring tools, the system empowers heritage custodians—from local museums to UNESCO World Heritage Sites—to adopt proactive conservation strategies. Its low-cost, open-source design aligns with the principles of Open Science, fostering collaboration and knowledge sharing among researchers and practitioners (Vicente-Saez and Martinez-Fuentes 2018). Moreover, the system's modularity and scalability make it adaptable to diverse heritage contexts, from indoor museum collections to outdoor archaeological sites.

Future research should address the system's limitations through larger-scale deployments and field testing. For instance, integrating energy-harvesting technologies, such as solar panels or piezoelectric devices, could enhance its sustainability for remote locations (Callebaut et al. 2021). Additionally, exploring alternative communication protocols, such as LoRa or ZigBee, could improve connectivity in areas with limited Wi-Fi coverage (Zahmatkesh and Al-Turjman 2020). These advancements would further solidify the system's role as a transformative tool for heritage conservation.

# Conclusions

This study successfully demonstrated the feasibility of a low-cost, modular IoT system for environmental monitoring and visitor behaviour analysis in cultural heritage settings. The system's integration of ESP32-based sensor nodes, Raspberry Pi edge computing, and Azure cloud services achieved reliable performance in a simulated museum environment, recording 14,149 data points over a 24-hour testing period. Key findings include the system's ability to accurately monitor environmental variables (e.g., temperature, humidity, and visitor proximity) and classify visitor behaviour using deep learning models, with the fine-tuned Vision Transformer (ViT) achieving 95% accuracy for multimodal classification. The photoelectric switch and ultrasonic distance sensor provided 100% reliability in detecting unauthorised intrusions, highlighting the system's potential for artefact protection.

The system's modular design and open-source architecture offer significant advantages over proprietary solutions, enabling cost-effective customisation and scalability. For instance, the JSON-based configuration file (wsn_config.json) allows users to add or remove sensors without modifying firmware, while the Raspberry Pi hub supports dynamic task allocation for data aggregation and image analysis. These features make the system adaptable to diverse heritage contexts, from indoor museum collections to outdoor archaeological sites.

However, several challenges remain. The system's reliance on Wi-Fi for camera node communication and its power consumption (e.g., 126 mAh per hour for ESP32-CAM nodes) limit its suitability for remote or off-grid locations. Future work should explore energy-harvesting technologies, such as solar panels or piezoelectric devices, to enhance sustainability (Callebaut et al., 2021). Additionally, alternative communication protocols, such as LoRa or ZigBee, could improve connectivity in areas with limited internet infrastructure (Zahmatkesh and Al-Turjman, 2020).

The system's low-cost design (<£200 per node) and open-source framework democratise access to advanced monitoring tools, particularly for resource-constrained institutions in developing countries. By empowering heritage custodians to adopt proactive conservation strategies, this work contributes to the global effort to preserve cultural heritage for future generations. Future research should focus on larger-scale deployments and multi-site integration to validate the system's performance in real-world settings, ensuring its readiness for widespread adoption.

# References

Agbota, Henoc, John Mitchell, Marianne Odlyha, and Matija Strlič. 2014. 'Remote Assessment of Cultural Heritage Environments with Wireless Sensor Array Networks'. *Sensors* 14 (5): 8779–93. https://doi.org/10.3390/s140508779.

Bacco, Manlio, Paolo Barsocchi, Pietro Cassara, Danila Germanese, Alberto Gotta, Giuseppe Riccardo Leone, Davide Moroni, Maria Antonietta Pascali, and Marco Tampucci. 2020. 'Monitoring Ancient Buildings: Real Deployment of an IoT System Enhanced by UAVs and Virtual Reality'. *IEEE Access* 8:50131–48. https://doi.org/10.1109/ACCESS.2020.2980359.

Callebaut, Gilles, Guus Leenders, Jarne Van Mulders, Geoffrey Ottoy, Lieven De Strycker, and Liesbet Van Der Perre. 2021. 'The Art of Designing Remote IoT Devices—Technologies and Strategies for a Long Battery Life'. *Sensors* 21 (3): 913. https://doi.org/10.3390/s21030913.

Camuffo, Dario, Antonio Della Valle, and Francesca Becherini. 2022. 'The European Standard EN 15757 Concerning Specifications for Relative Humidity: Suggested Improvements for Its Revision'. *Atmosphere* 13 (9): 1344. https://doi.org/10.3390/atmos13091344.

Casillo, Mario, Francesco Colace, Brij B. Gupta, Angelo Lorusso, Francesco Marongiu, and Domenico Santaniello. 2022. 'A Deep Learning Approach to Protecting Cultural Heritage Buildings Through IoT-Based Systems'. In *2022 IEEE International Conference on Smart Computing (SMARTCOMP)*, 252–56. Helsinki, Finland: IEEE. https://doi.org/10.1109/SMARTCOMP55677.2022.00063.

Chen, Jiasi, and Xukan Ran. 2019. 'Deep Learning With Edge Computing: A Review'. *Proceedings of the IEEE* 107 (8): 1655–74. https://doi.org/10.1109/JPROC.2019.2921977.

Dian, F. John, Amirhossein Yousefi, and Sungjoon Lim. 2018. 'A Practical Study on Bluetooth Low Energy (BLE) Throughput'. In *2018 IEEE 9th Annual Information Technology, Electronics and Mobile Communication Conference (IEMCON)*, 768–71. Vancouver, BC: IEEE. https://doi.org/10.1109/IEMCON.2018.8614763.

Donders, Yvonne. 2020. 'Cultural Heritage and Human Rights'. In *Oxford Handbook on International Cultural Heritage Law*, 1–29. Oxford University Press / Amsterdam Center for International Law. https://papers.ssrn.com/sol3/papers.cfm?abstract_id=3642413.

Grassini, Sabrina, Emma Angelini, Ahmed Elsayed, Simone Corbellini, Luca Lombardo, and Marco Parvis. 2017. 'Cloud Infrastructure for Museum Environmental Monitoring'. In *2017 IEEE International Instrumentation and Measurement Technology Conference (I2MTC)*, 1–6. Torino, Italy: IEEE. https://doi.org/10.1109/I2MTC.2017.7969984.

Haugen, Annika, Chiara Bertolin, Gustaf Leijonhufvud, Tone Olstad, and Tor Broström. 2018. 'A Methodology for Long-Term Monitoring of Climate Change Impacts on Historic Buildings'. *Geosciences* 8 (10): 370. https://doi.org/10.3390/geosciences8100370.

Krommyda, Maria, Nikos Mitro, and Angelos Amditis. 2022. 'Smart IoT Sensor Network for Monitoring of Cultural Heritage Monuments'. In *Smart Trends in Computing and Communications*, edited by Yu-Dong Zhang, Tomonobu Senjyu, Chakchai So-In, and Amit Joshi, 286:175–84. Lecture Notes in Networks and Systems. Singapore: Springer Singapore. https://doi.org/10.1007/978-981-16-4016-2_17.

Kumar, Pakhee, Ferda Ofli, Muhammad Imran, and Carlos Castillo. 2020. 'Detection of Disaster-Affected Cultural Heritage Sites from Social Media Images Using Deep Learning Techniques'. *Journal on Computing and Cultural Heritage* 13 (3): 1–31. https://doi.org/10.1145/3383314.

Leccese, Fabio, Marco Cagnetti, Andrea Calogero, Daniele Trinca, Stefano Pasquale, Sabino Giarnetti, and Lorenzo Cozzella. 2014. 'A New Acquisition and Imaging System for Environmental Measurements: An Experience on the Italian Cultural Heritage'. *Sensors* 14 (5): 9290–9312. https://doi.org/10.3390/s140509290.

Lee, Woosik, and Dong-hoon Lee. 2019. 'Cultural Heritage and the Intelligent Internet of Things'. *Journal on Computing and Cultural Heritage* 12 (3): 1–14. https://doi.org/10.1145/3316414.

Manfriani, Chiara, Giovanni Gualdani, Giacomo Goli, Bruce Carlson, Anna Rita Certo, Paola Mazzanti, and Marco Fioravanti. 2021. 'The Contribution of IoT to the Implementation of Preventive Conservation According to European Standards: The Case Study of the "Cannone" Violin and Its Historical Copy'. *Sustainability* 13 (4): 1900. https://doi.org/10.3390/su13041900.

Mansouri, Yaser, and Rajkumar Buyya. 2020. 'Data Access Management System in Azure Blob Storage and AWS S3 Multi-Cloud Storage Environments': In *Advances in Information Security, Privacy, and Ethics*, edited by Brij B. Gupta and Srivathsan Srinivasagopalan, 130–47. IGI Global. https://doi.org/10.4018/978-1-7998-2242-4.ch007.

Meha, Arbresha, Alberta Tahiri, and Mimoza Zhubi. 2020. 'The Importance of Cultural Heritage in Tourism Development-the Case of Kosovo'. *Acta Universitatis Danubius. Œconomica* 16 (6).

Mesas-Carrascosa, Francisco, Daniel Verdú Santano, Jose Meroño De Larriva, Rafael Ortíz Cordero, Rafael Hidalgo Fernández, and Alfonso García-Ferrer. 2016. 'Monitoring Heritage Buildings with Open Source Hardware Sensors: A Case Study of the Mosque-Cathedral of Córdoba'. *Sensors* 16 (10): 1620. https://doi.org/10.3390/s16101620.

Mitro, Nikos, Maria Krommyda, and Angelos Amditis. 2022. 'Smart Tags: IoT Sensors for Monitoring the Micro-Climate of Cultural Heritage Monuments'. *Applied Sciences* 12 (5): 2315. https://doi.org/10.3390/app12052315.

Monti, Giorgio, Fabio Fumagalli, Giuseppe Quaranta, Marco Sgroi, and Marcello Tommasi. 2018. 'A Permanent Wireless Dynamic Monitoring System for the Colosseum in Rome'. *Journal of Structural Integrity and Maintenance* 3 (2): 75–85. https://doi.org/10.1080/24705314.2018.1463020.

Piccialli, Francesco, Salvatore Cuomo, Vincenzo Schiano Di Cola, and Giampaolo Casolla. 2024. 'A Machine Learning Approach for IoT Cultural Data'. *Journal of Ambient Intelligence and Humanized Computing* 15 (2): 1715–26. https://doi.org/10.1007/s12652-019-01452-6.

Shi, Weisong, Jie Cao, Quan Zhang, Youhuizi Li, and Lanyu Xu. 2016. 'Edge Computing: Vision and Challenges'. *IEEE Internet of Things Journal* 3 (5): 637–46. https://doi.org/10.1109/JIOT.2016.2579198.

Silva, Hugo Entradas, Guilherme B.A. Coelho, and Fernando M.A. Henriques. 2020. 'Climate Monitoring in World Heritage List Buildings with Low-Cost Data Loggers: The Case of the Jerónimos Monastery in Lisbon (Portugal)'. *Journal of Building Engineering* 28 (March):101029. https://doi.org/10.1016/j.jobe.2019.101029.

Sudantha, BH, KMHK Warnakulasooriya, YP Jayasuriya, GR Ratnayaka, PKS Mahanama, EJ Warusavitharana, and SN Weerasinghe. 2018. 'Open-Source Implementation of an Integrated, Low-Cost Environmental Monitoring System (EMS) for Developing Countries'. *Bhumi, The Planning Research Journal* 6 (1).

Vicente-Saez, Ruben, and Clara Martinez-Fuentes. 2018. 'Open Science Now: A Systematic Literature Review for an Integrated Definition'. *Journal of Business Research* 88 (July):428–36. https://doi.org/10.1016/j.jbusres.2017.12.043.

Vidovszky, I, and F Pintér. 2022. 'IoT System for Maintenance Monitoring of Historic Buildings – Smart Monitoring'. *IOP Conference Series: Materials Science and Engineering* 1218 (1): 012007. https://doi.org/10.1088/1757-899X/1218/1/012007.

Zahmatkesh, Hadi, and Fadi Al-Turjman. 2020. 'A Review on the Use of Wireless Sensor Networks in Cultural Heritage: Communication Technologies, Requirements, and Challenges'. In *Trends* *in Cloud-Based IoT*, edited by Fadi Al-Turjman. EAI/Springer Innovations in Communication and Computing. Cham: Springer International Publishing. https://doi.org/10.1007/978-3-030-40037-8.

Zhang, Jingwen, Meiting Guo, Biyan Li, and Ruimin Lu. 2021. 'A Transport Monitoring System for Cultural Relics Protection Based on Blockchain and Internet of Things'. *Journal of Cultural Heritage* 50 (July):106–14. https://doi.org/10.1016/j.culher.2021.05.007.

Zombory, Katarzyna. 2022. 'The Protection of Cultural Heritage in International Law'. In *Legal Studies on Central Europe*, edited by Anikó Raisz, 9:239–62. Central European Academic Publishing. https://doi.org/10.54171/2022.ar.ilfcec_11.

| Study case reference | Location | Type of asset | Sensors used | Real time visualisation | Deep learning | Use of images | Use of cloud | Open Hardware / Source | IoT |
|---|---|---|---|---|---|---|---|---|---|
| (Casillo *et al.*, 2022) | Pompeii, Italy | Building | T, RH, WS | Yes | Yes | Yes | Yes | Unknown | Yes |
| (Nedjah *et al.*, 2022, pp. 149–152) | Herculaneum, Italy | Building | T, RH, C, SH | Yes | No | Yes | Yes | Unknown | Yes |
| (Vidovszky and Pintér, 2022) | PoC | Building | T, RH, C | Yes | No | Yes | Yes | Yes | Yes |
| (Krommyda, Mitro and Amditis, 2022, pp. 175–184) | PoC | Building | T, RH, SH | Yes | No | No | Yes | Yes | Yes |
| (Manfriani *et al.*, 2021) | Genoa, Italy | Museum collections | T, RH | No | No | No | No | No | No |
| (Zhang *et al.*, 2021) | Guangdong, China | Museum collections | T, RH, V, C | Yes | No | Yes | Yes | Unknown | Yes |
| (Silva, Coelho and Henriques, 2020) | Lisbon, Portugal | Building | AQ, T, RH | No | No | No | No | Yes | No |
| (Bacco *et al.*, 2020) | Leghorn, Italy | Building | Ac, T, RH, WS | Yes | Yes | Yes | Yes | Unknown | Yes |
| (Maksimovic and Cosovic, 2019) | Sarajevo, Bosnia, and Herzegovina | Building | AQ, RH, T, V | Yes | Yes | No | Yes | Yes | Yes |
| (Monti *et al.*, 2018) | Rome, Italy | Building | V | Yes | No | No | No | No | No |
| (Mesas-Carrascosa *et al.*, 2016) | Cordoba, Spain | Building | RH, T | No | No | No | No | Yes | No |
| (Varas-Muriel, Martínez-Garrido and Fort, 2014) | Madrid, Spain | Building | RH, T | No | No | No | No | No | No |
| (Agbota *et al.*, 2014) | London, UK | Building | AQ | Yes | No | No | Yes | Yes | Yes |
| (Prasanth *et al.*, 2021) | PoC | Museum collections | RH, T, L, M | Yes | No | No | Yes | No | Yes |

*Table 1: Examples of environment monitoring systems for Cultural Heritage Assets. Abbreviations used: PoC: Proof of Concept, T: Temperature, RH: Relative Humidity, WS: Weather sensors, C: Camera, TC: Thermal Camera, SH: Structure Humidity, V: Vibration*

| Node | MCU | Sensor | Power consumption 1h | Communication script runtime | RSSI at 1m |
|---|---|---|---|---|---|
| **Temperature and humidity node** | ESP32-WROOM-32 | DHT22 | 84 mAh | 23 seconds per measurement (46 for both values) | -90 dBm |
| **Presence sensor node** | ESP32-WROOM-32 | HW-416-B | 114 mAh | 12 seconds | -101 dBm |
| **Photo-electric switch sensor node** | ESP32-WROOM-32 | E3F-R2N1 | 54 mAh | 13 seconds | -86 dBm |
| **Sound sensor node** | ESP32-WROOM-32 | KY-038 | 60 mAh | 11 seconds | -89 dBm |
| **Distance sensor node** | ESP32-WROOM-32 | HC-SR04 | 54 mAh | 6 seconds | -80 dBm |
| **Camera node** | ESP32-S | OV2640 camera | 120-126 mAh | 4 seconds | NA |

*Table 2: Senor nodes of the WSN*

| Deep learning strategy | Model | Multimodal approach | | Binary approach | |
|---|---|---|---|---|---|
| | | Training time (80% dataset) | Accuracy | Training time (80% dataset) | Accuracy |
| Bespoken ANNs | CNN | **4 minutes 9 seconds** | 0.93 | 2 minutes 9 seconds | 0.84 |
| | FNN | 2 mins 18 seconds | 0.83 | 1 minute 21 seconds | 0.80 |
| | RNN | **23 seconds** | 0.69 | 12 seconds | 0.85 |
| Fine tunned ANN | google-ViT | 14 mins 47 seconds | 0.95 | 7 mins 50 seconds | 1.0 |
| Zero Shot classifier | CLIP | NA | 0.19 | NA | 0.64 |

*Table 3: Performance of the different deep learning approaches*

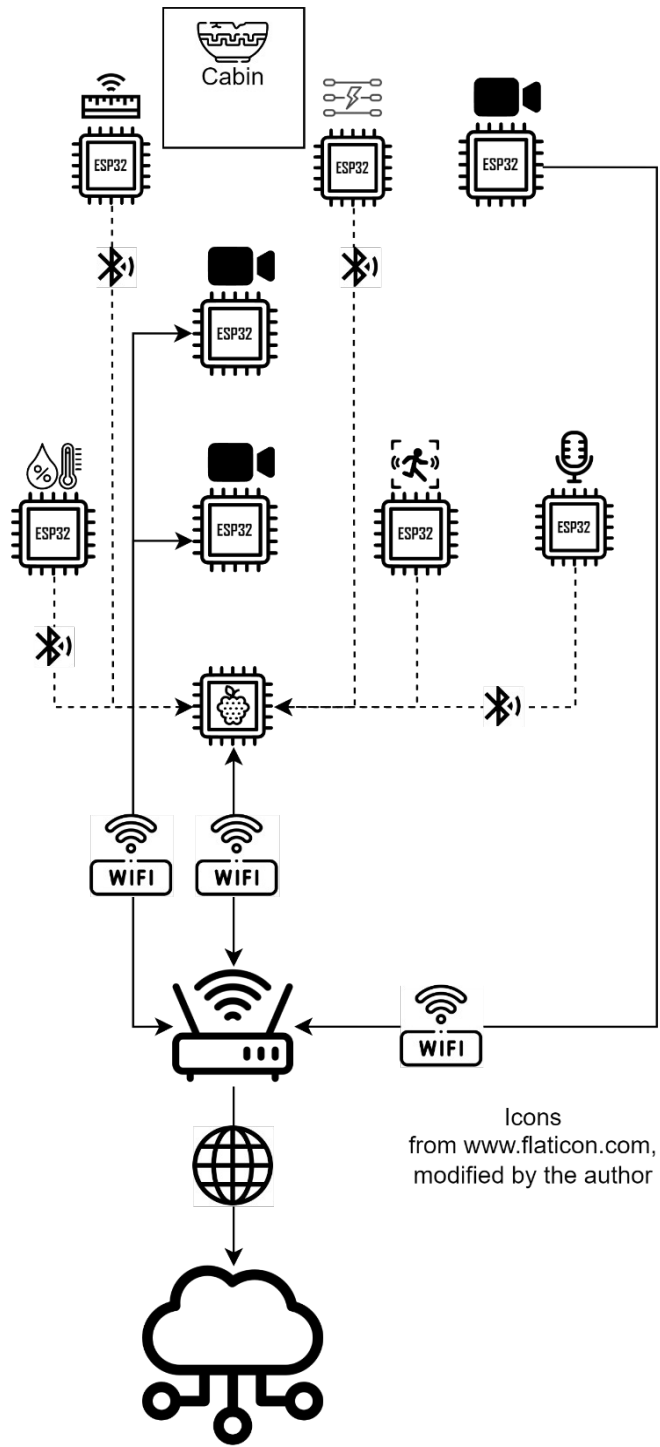

*Figure 1: Diagram of the WSN and CH*

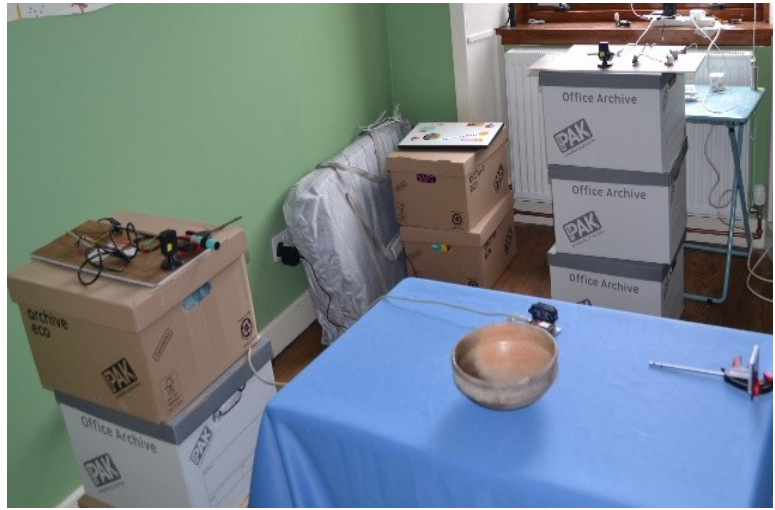

*Figure 2: Experimental set up, general view*

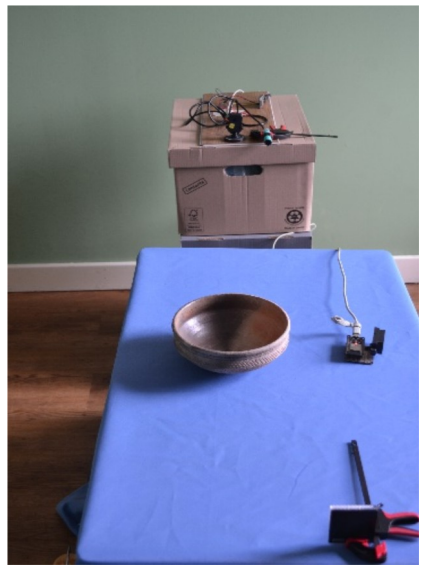

*Figure 3: Experimental set up, detail of photoelectric switch, distance sensor and camera module*

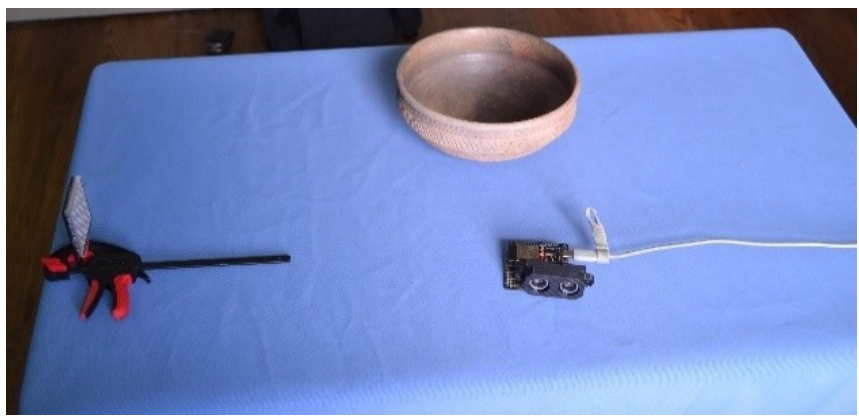

*Figure 4: Experimental set up, detail of distance sensor*

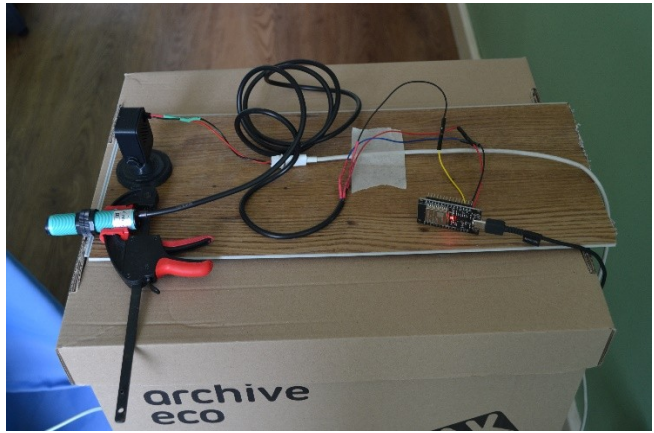

*Figure 5: Experimental set up, detail of photoelectric switch and camera module*

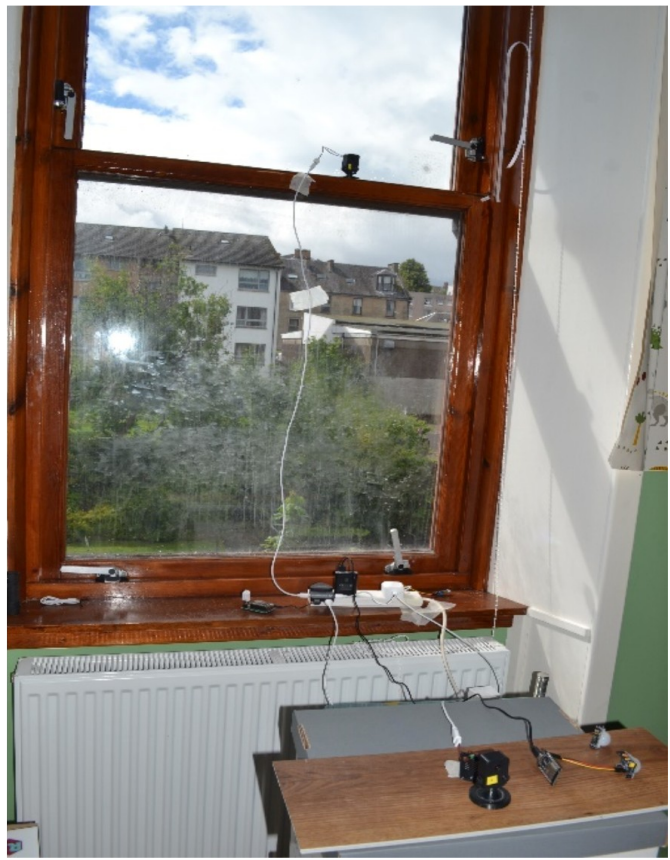

*Figure 6: Experimental set up, detail of camera modules, PIR, sound and temperature and humidity sensors*

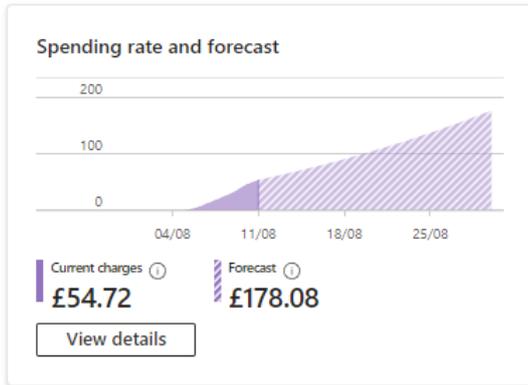

*Figure 7: Spending analysis of the cloud infrastructure*